\documentclass[conference]{IEEEtran}
\IEEEoverridecommandlockouts

\usepackage{amsmath,amssymb,amsfonts}
\usepackage{algorithmic}
\usepackage{graphicx}
\usepackage{textcomp}
\usepackage{xcolor}

\def\BibTeX{{\rm B\kern-.05em{\sc i\kern-.025em b}\kern-.08em
    T\kern-.1667em\lower.7ex\hbox{E}\kern-.125emX}}

\def\orcid#1{\kern.08em\href{https://orcid.org/#1}{\protect\includegraphics[keepaspectratio,width=0.7em]{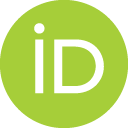}}}

\usepackage{circledsteps}
\usepackage{circledtext}
\usepackage[nolist]{acronym}
\usepackage{xspace}
\usepackage{array}
\usepackage[subrefformat=parens,labelformat=parens]{subcaption}
\usepackage{etoolbox}
\usepackage{csquotes}
\usepackage{booktabs}
\usepackage{multirow}
\usepackage{numprint}
\usepackage{microtype}
\usepackage{float}
\usepackage{balance}
\usepackage{hyperref}

\usepackage[style=numeric,maxbibnames=5,backend=biber,sortcites=true]{biblatex}
\AtEveryBibitem{%
  \clearfield{doi}
  \clearfield{biburl}
  \clearname{editor}
  \clearfield{url}
}

\usepackage[]{mdframed}

\pgfkeys{/csteps/inner color=white}
\pgfkeys{/csteps/fill color=black}

\circledtextset{boxfill=black}
\circledtextset{boxtype=o}
\circledtextset{charf=\bfseries\Large}
\circledtextset{charcolor=white}
\circledtextset{charshrink=0.6}
\circledtextset{resize=real}
\circledtextset{width=0.8em}

\newcommand{\etal}{et~al.\xspace}
\newcommand{\eg}{e.g.\xspace}
\newcommand{\ie}{i.e.\xspace}

\newcommand{\trojan}{trojan\xspace}

\newcommand{\trojans}{trojans\xspace}

\usepackage{tikz}
\usepackage{pgfplots}
\pgfplotsset{compat=1.8}
\usetikzlibrary{calc}
\usetikzlibrary{positioning,arrows,fit,matrix}
\usepgfplotslibrary{groupplots}
\usepgfplotslibrary{fillbetween}
\usepgfplotslibrary{statistics}

\usepackage{tabularx}
\newcolumntype{Y}{>{\centering\arraybackslash}X}
\newcolumntype{L}[1]{>{\raggedright\let\newline\\\arraybackslash\hspace{0pt}}m{#1}}
\newcolumntype{C}[1]{>{\centering\let\newline\\\arraybackslash\hspace{0pt}}m{#1}}
\newcolumntype{R}[1]{>{\raggedleft\let\newline\\\arraybackslash\hspace{0pt}}m{#1}}

\newcolumntype{P}[1]{>{\centering\arraybackslash}p{#1}}

\DeclareMathOperator*{\argmin}{arg\,min}

\begin{acronym}
    \acro{AI}{artificial intelligence}
    \acro{ALU}{arithmetic logic unit}
    \acro{API}{application programming interface}
    \acro{APU}{application processing unit}
    \acro{ASIC}{application-specific integrated circuit}
    \acro{AXI}{advanced extensible interface}

    \acro{BRAM}{block \acs{RAM}}
    
    \acro{CPU}{central processing unit}
    \acro{CRT}{chinese remainder theorem}
    
    \acro{DL}{deep learning}
    \acro{DLL}{dynamic link library}
    \acroplural{DLL}[DLLs]{dynamic link libraries}
    \acro{DNN}{deep neural network}
    \acro{DPU}{deep learning processing unit}
    \acro{DRM}{digital rights management}
    \acro{DSP}{digital signal processor}
    \acro{DSR}{desired success rate}
    
    \acro{ECC}{elliptic-curve cryptography}
    \acro{ECU}{electronic control unit}
    \acro{EDA}{electronic design automation}

    \acro{FF}{flip-flop}
    \acro{FPGA}{field-programmable gate array}
    \acro{FSM}{finite state machine}
    
    \acro{GCD}{greatest common divisor}
    \acro{GCD}{greatest common divisor}
    \acro{GPU}{graphics processing unit}
    
    \acro{HDL}{hardware description language}
    
    \acro{IC}{integrated circuit}
    \acro{IEEE}{institute of electrical and electronics engineers}
    \acro{IP}{intellectual property}

    \acro{LUT}{lookup table}
    
    \acro{MATE}{man-at-the-end}
    \acro{MITM}{man-in-the-middle}
    \acro{ML}{machine learning}
    \acro{MUX}{multiplexer}

    \acro{OEM}{original equipment manufacturer}

    \acro{PDK}{process design kit}
    \acro{PL}{programmable logic}
    \acro{PKI}{public key infrastructure}
    \acro{PS}{processing system}
    
    \acro{RAM}{random-access memory}
    \acro{ROM}{read-only memory}
    \acro{RTL}{register transfer level}

    \acro{SEM}{scanning electron microscope}
    \acro{SGD}{stochastic gradient descent}
    \acro{SoC}{system-on-chip}
    \acroplural{SoC}[SoCs]{systems-on-chip}

    \acro{XIR}{Xilinx intermediate representation}

\end{acronym}

\begin{document}

\title{Evil from Within: Machine Learning Backdoors\\Through Dormant Hardware Trojans}

\author{\IEEEauthorblockN{
Alexander Warnecke\IEEEauthorrefmark{1}\IEEEauthorrefmark{2}\IEEEauthorrefmark{3}\thanks{\IEEEauthorrefmark{1}Both authors contributed equally.}\orcid{0009-0006-3617-3968},
Julian Speith\IEEEauthorrefmark{1}\IEEEauthorrefmark{4}\orcid{0000-0002-8408-8518},
Jan-Niklas Möller\IEEEauthorrefmark{4}\orcid{0009-0007-3006-7846},
Konrad Rieck\IEEEauthorrefmark{2}\IEEEauthorrefmark{3}\orcid{0000-0002-5054-8758},
Christof Paar\IEEEauthorrefmark{4}\orcid{0000-0001-8681-2277}}
\IEEEauthorblockA{
\IEEEauthorrefmark{2}Berlin Institute for the Foundations of Learning and Data (BIFOLD)\\
\IEEEauthorrefmark{3}Technische Universität Berlin\\
\IEEEauthorrefmark{4}Max Planck Institute for Security and Privacy (MPI-SP)}
}

\maketitle

\begin{abstract}
Backdoors pose a severe threat to machine learning, as they can compromise the integrity of security-critical systems, such as self-driving cars.
While different defenses have been proposed to address this threat, they all rely on the assumption that the hardware accelerator executing a learning model is trusted.
This paper challenges this assumption and investigates a backdoor attack that completely resides within such an accelerator.
Outside of the hardware, neither the learning model nor the software is manipulated so that current defenses fail.
As memory on a hardware accelerator is limited, we utilize \emph{minimal backdoors} that deviate from the original model by a few model parameters only.
To mount the backdoor, we develop a \emph{hardware trojan} that lays dormant until it is programmed after in-field deployment. 
The trojan can be provisioned with the minimal backdoor and performs a parameter replacement only when the target model is processed.
We demonstrate the feasibility of our attack by implanting our hardware trojan into a commercial machine-learning accelerator and programming it with a minimal backdoor for a traffic-sign recognition system. 
The backdoor affects only 30 model parameters (0.069\%) with a backdoor trigger covering 6.25\% of the input image, yet it reliably manipulates the recognition once the input contains a backdoor trigger. 
Our attack expands the circuit size of the accelerator by only 0.24\% and does not increase the run-time, rendering detection hardly possible.
Given the distributed hardware manufacturing process, our work points to a new threat in machine learning that currently eludes security mechanisms.
\end{abstract}


\section{Introduction}\label{dpu_trojan::sec::intro}
Machine learning has become ubiquitous in recent years, with applications ranging from traffic sign recognition~\cite{DBLP:journals/ivc/EscaleraAM03} over cancer detection~\cite{DBLP:journals/nature/EstevaKNKSBT17} and protein folding~\cite{jumper2021highly} 
to numerous use cases in social networks~\cite{DBLP:conf/hpca/WuBCCCDHIJJLLLQ19,DBLP:conf/hpca/HazelwoodBBCDDF18}.
This development was driven by advances in hardware acceleration, allowing complex learning models, such as deep neural networks, to run even on systems with limited resources. 
Today, hardware acceleration is indispensable in many systems that use machine learning.
The adoption of machine learning in practice is overshadowed by attacks that range from adversarial examples to backdoors and poisoning~\cite{DBLP:conf/icml/BiggioNL12,DBLP:conf/sp/Carlini017, DBLP:conf/eurosp/PapernotMSW18}. 
Previous work has explored these threats and developed defenses of varying robustness~\cite{DBLP:conf/sp/WangYSLVZZ19,DBLP:conf/sp/XuWLBGL21,DBLP:conf/nips/Tran0M18,DBLP:conf/acsac/GaoXW0RN19}. 
A key assumption is that the hardware running the learning models is trustworthy. 
That is, ensuring the integrity of the input and the learning model to realize secure machine-learning applications in practice is deemed sufficient.

In this paper, we challenge this assumption.
Hardware manufacturing is far from transparent, involving opaque components and untrusted parties.
A multitude of attack vectors arise from the design process of \acp{IC} alone~\cite{DBLP:conf/fdtc/BhasinDGNS13, DBLP:journals/dt/TehranipoorK10a, DBLP:journals/computer/KarriRRT10} and their use of third-party \ac{IP}~\cite{DBLP:conf/fdtc/BhasinDGNS13, DBLP:journals/todaes/XiaoFJKBT16}. 
Given the complexity of modern circuits, built from billions of nanometer-sized transistors, it is difficult (if not impossible) to verify that an \ac{IC} provides the exact logic specified in its design.
In fact, this problem has led governments to enforce control over the hardware supply chain and subsidize domestic manufacturing, \eg, through the \emph{European Chips Act}~\cite{euchips2022} and the \emph{US CHIPS and Science Act}~\cite{uschips2022}.

We exploit this opacity of hardware and explore the design space of a backdoor attack that entirely resides within the hardware of a machine-learning accelerator.
Thereby, we investigate the threat potential posed by hardware trojans to machine-learning acceleration. 
To mount a targeted backdoor attack---be it in hardware or also just in the general case---an attacker must know the executed learning model.
However, machine-learning accelerators such as Google's TPU and Apple's Neural Engine are designed and manufactured independent of the exact learning models they later execute.
Usually, the trained learning model does not even exist when the hardware accelerator is being built and often the manufacturer of the accelerator and the provider of the model are distinct entities.
Therefore, we must assume that the learning model is unknown when a hardware trojan is inserted during hardware design or manufacturing.
Hence, this setting demands a programmable hardware trojan design that can be updated after in-field deployment.

\begin{figure}[htb]
    \centering
    \includegraphics[width=0.95\linewidth]{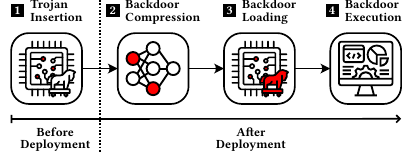}
    \caption{Overview of our hardware-based backdoor attack.}
    \label{dpu_trojan::fig::attack_flow_overview}
\end{figure}

Against this background, we propose a hardware trojan attack that, during inference, selectively replaces model parameters in the hardware. 
Outside the hardware, the learning model remains unchanged; thus, defenses operating on the model itself inevitably fail. 
\autoref{dpu_trojan::fig::attack_flow_overview} shows our four-stage attack.
We use a traffic sign recognition system of a self-driving car as a running example.

First~\circledtext[boxtype=O]{1}, a dormant, programmable hardware trojan is inserted into a hardware accelerator. 
Potential attackers range from the designer to a malicious supplier, which are common threat models in hardware trojan research~\cite{DBLP:conf/ches/SkorobogatovW12,DBLP:conf/sp/SturtonHWK11,DBLP:conf/ches/BeckerRPB13,DBLP:conf/sp/PuschnerMBKMP23}.
At this stage, the trojan is still dormant and does not yet affect inference.
After deployment of the accelerator~\circledtext[boxtype=O]{2}, the adversary obtains the learning model and computes a minimal backdoor that induces a misclassification whenever a certain trigger pattern is present in the input.
This stage is performed after hardware manufacturing, for example, by extracting a model in-field~\cite{DBLP:conf/uss/SunSLM21}.
Next~\circledtext[boxtype=O]{3}, the adversary programs the trojan with the backdoor. 
This can be done via over-the-air updates or by manipulating the car directly, \eg, in a workshop.
Finally~\circledtext[boxtype=O]{4}, the targeted model is executed on the accelerator, generating incorrect predictions only if the backdoor trigger is present.

\subsection{Related Work and Open Challenges}

\par\smallskip\textbf{Machine-Learning Backdoors.}
A machine learning backdoor is a covertly implanted vulnerability in a model's architecture, designed to trigger specific behaviors or outputs when activated by a predefined trigger signal, often leading to malicious or unintended consequences.
Gu~\etal~\cite{DBLP:journals/access/GuLDG19} show that an attacker can implant a backdoor by injecting malicious samples into the training dataset. 
Further approaches relax the assumption of access to the training data~\cite{DBLP:conf/ndss/LiuMALZW018}, the visibility and position of the trigger~\cite{DBLP:conf/eurosp/SalemWBMZ22, DBLP:conf/codaspy/ZhongLSZ020,DBLP:conf/aaai/SahaSP20,DBLP:conf/nips/NguyenT20}, or the number of malicious examples required~\cite{DBLP:conf/nips/ShafahiHNSSDG18}. 
Tang~\etal~\cite{DBLP:conf/kdd/TangDLYH20} assume that an attacker can insert additional neuron connections to implant the backdoor. 
Stealthy backdoors that are inserted during model compilation~\cite{DBLP:conf/satml/CliffordSZAM24}, model quantization~\cite{DBLP:journals/tdsc/MaQGZAXFZAA24}, or implemented by the software execution environment~\cite{DBLP:conf/icse/LiH0CL21} were also proposed.

Hardware acceleration comes with the unique challenge of being unable to hold an entire learning model in the hardware at once due to memory limitations.
As we place our trojan in the hardware accelerator, we are constrained in the amount of memory available for storing the parameters of the backdoored model.
Hence, our trojan can always only store a select few parameters.
Multiple attacks are optimized towards creating backdoors with very few parameter changes compared to the original model~\cite{DBLP:conf/cvpr/RakinHF20,DBLP:conf/dsn/TolIASZ23}. 
Still, they only work for one specific set of parameters and can lead to drops in test accuracy of up to $15\%$~\cite{DBLP:conf/cvpr/RakinHF20}. 
Therefore, we need a new approach that minimizes the number of parameter changes, enables an effective attack, achieves good classification performance, and still succeeds in the presence of small parameter changes, \eg, when the learning model is fine-tuned with new data.
This task is complicated by \emph{quantization} of the model parameters, which maps floating point parameters to a narrow bit width.
Quantization is frequently performed for hardware acceleration~\cite{DBLP:conf/aaai/ZhouMCF18,DBLP:conf/cvpr/WangLLLH19}. 
Therefore, we must balance the number of changed parameters and their amplitude.

\begin{mdframed}
    \textbf{Challenge 1: Minimal Backdoor.}
    How can we design a backdoor that changes as few parameters as possible while maintaining a high classification score?
    Can such a backdoor be robust even in the presence of quantization?
\end{mdframed}

\par\smallskip\textbf{Hardware \trojans and Fault Injection Attacks.}
For an overview of hardware trojans, we refer the interested reader to the many summaries in the area~\cite{DBLP:journals/iet-cdt/XueGLYO20,DBLP:journals/integration/LiLZ16,DBLP:journals/dt/TehranipoorK10a}.
The idea of hardware trojans targeting neural networks was first proposed by Clements~\etal~\cite{DBLP:journals/corr/abs-1806-05768} and Li~\etal~\cite{DBLP:conf/isvlsi/LiYNWWWY18}. 
Other works~\cite{DBLP:conf/ats/YeH018} require manipulations to the inputs to trigger the hardware trojan which then bypasses the machine-learning accelerator altogether.
More recent \trojan attacks trigger on intermediate layer outputs~\cite{DBLP:journals/corr/abs-1911-00783}, are inserted into the on-chip memory controller~\cite{DBLP:journals/tcad/HuZDLZYLX21}, or target activation parameters~\cite{DBLP:journals/esl/MukherjeeC22} for accuracy degradation. 
Liu~\etal~\cite{DBLP:conf/dac/LiuCZL20} inject glitches for untargeted misclassification and demonstrate applicability using Xilinx Vitis AI. 
Many attacks assume that the model executed on the hardware is known during manufacturing; others require changes to the input images or are generally inflexible when it comes to model updates. 
However, a hardware accelerator is designed model-agnosticically and can be equipped with various learning models after shipment.

A related line of research deals with fault injection attacks that aim to compromise learning models by changing their parameters in memory, for example by flipping single bits. 
Various works perform physical attacks~\cite{DBLP:conf/iccad/LiuWLX17,DBLP:conf/ccs/BreierHJMB018, DBLP:conf/uss/HongFKGD19,DBLP:conf/fdtc/AlamTGTF19} and, for example, induce the bit flips by a laser in a lab environment. 
Building on the Rowhammer attack~\cite{DBLP:conf/isca/KimDKFLLWLM14}, multiple methods were proposed to find the bits most suitable for an attacker to flip for inserting backdors~\cite{DBLP:conf/cvpr/RakinHF20, DBLP:conf/dsn/TolIASZ23,DBLP:conf/dsn/TolIASZ23} or causing severe performance drops~\cite{DBLP:conf/iccv/RakinHF19,DBLP:conf/uss/YaoRF20,DBLP:conf/iclr/BaiWZL0X21}. 
Rowhammer attacks, however, require the attacker to have memory access and are unreliable in practice. 
Gruss~\etal~\cite{DBLP:conf/sp/GrussLSGJOSY18} show that a single bit flip could take days to accomplish while, at the same time, Yao~\etal\cite{DBLP:conf/uss/YaoRF20} find that multiple bit flips are required to attack a modern quantized neural network.
Furthermore, proactive defenses against such bit flips have recently been proposed~\cite{DBLP:conf/uss/0017YWYS23}.

\begin{mdframed}
    \textbf{Challenge 2: Hardware-based Attack.}
    How can we hide a machine-learning backdoor in hardware so it cannot be observed from the outside?
    Can such a backdoor be reliable and flexible enough to target any model?
\end{mdframed}

\par\smallskip\textbf{Countermeasures and Defenses.}
The presence of neural backdoors also spawned research on detection and defense mechanisms. 
One line of research tries to detect whether a trigger is present in the model, for example by finding shortcuts between output classes~\cite{DBLP:conf/sp/WangYSLVZZ19}, training meta models to classify networks~\cite{DBLP:conf/sp/XuWLBGL21}, or utilizing statistical properties from model predictions~\cite{DBLP:conf/ijcai/ChenFZK19,DBLP:conf/uss/Tang0TZ21}.
An orthogonal line of research tries to detect whether a given input image contains a trigger, \eg, by finding anomalies in activations or latent representations when propagating the input through the model~\cite{DBLP:conf/aaai/ChenCBLELMS19,DBLP:conf/nips/Tran0M18,DBLP:conf/acsac/GaoXW0RN19}.
Since our backdoor is only observable within the hardware accelerator, such countermeasures are evaded by our attack.

In hardware, trojan attacks can be detected by comparison with a trojan-free circuit~\cite{DBLP:conf/iscas/BhasinR15,DBLP:conf/sp/PuschnerMBKMP23}.
However, no such golden model exists in our settings as the designer or a malicious supplier inserts the trojan.
Even formal verification approaches~\cite{DBLP:conf/iscas/RathmairSK14,DBLP:conf/dac/GuoDJFM15} are ineffective as they would have to be performed by the malicious entity.
Similar arguments can be made for proof-carrying hardware~\cite{DBLP:journals/tifs/LoveJM12}, which additionally suffers from scaling issues~\cite{DBLP:conf/itc/NahiyanSVCFT17}.
Techniques such as information flow security verification require at least some knowledge of the \ac{IP} internals to identify \emph{observe points}~\cite{DBLP:conf/itc/NahiyanSVCFT17}.
The only viable option is to analyze the circuit for malicious functionality through reverse engineering, which is challenging on its own.

Still, the tampered hardware accelerator must perform its regular operation without any noticeable deviations to avoid raising suspicion. 
Since hardware accelerators for machine-learning are usually stateless and do not know the context in which they operate~\cite{DBLP:conf/hpec/ReutherM0GSK19,DBLP:journals/tcad/WangGYLXZ17}, a hardware trojan must decide for itself which parameters to replace during inference. 
At the same time, the attack overhead must remain low so that the critical path is not extended and no anomalous timings can be observed. 
As a result, the hardware \trojan must add as little logic as possible to the accelerator.

\begin{mdframed}
    \textbf{Challenge 3: Unobtrusive Operation.}
    How can a hardware trojan inject an effective, targeted backdoor within a stateless accelerator without raising suspicion?
\end{mdframed}

\subsection{Contributions}
By overcoming these challenges, we demonstrate the practical feasibility of hardware trojan attacks on machine-learning models in a real-world setting.
We make the following contributions:
\begin{itemize}
    \setlength{\itemsep}{3pt}
    \item \textbf{Hardware Trojan.} 
    We explore the design space for a programmable hardware trojan that injects a backdoor into a learning model upon inference on a hardware accelerator. 
    To this end, we propose a novel trojan design that can be programmed independently of the hardware manufacturing process, see~\autoref{dpu_trojan::sec::trojan_framework}. 
    
    \item \textbf{Minimal Backdoors.} 
    We expand on the concept of minimal backdoors for machine-learning models in the context of hardware acceleration. 
    These backdoors are optimized to change as few parameters as possible while maintaining prediction accuracy to comply with memory limitations of the hardware platform and remain stealthy, see~\autoref{dpu_trojan::sec::ml_trojan}.

    \item \textbf{Real-World Case Study.} 
    We show the feasibility of our attack by trojanizing a commercial \ac{IP} core for machine-learning acceleration, \ie, the Xilinx Vitis AI \acs{DPU}.
    Our \trojan causes stop signs to be interpreted as right-of-way, potentially with fatal consequences if deployed in the real world.
    Despite replacing only $0.069\%$ of the parameters, the backdoor is reliably activated by a trigger that covers only $6.25\%$ of the input image, see~\autoref{dpu_trojan::sec::case_study}.
\end{itemize}

\section{Backdoor Attack Overview}
\label{dpu_trojan::sec::trojan_framework}
In the following section we provide an overview of our backdoor attack by formalizing the underlying attacker model. To this end, we continue with our running example of backdooring a traffic-sign recognition model during execution.
\subsection{Attacker Model}
\label{dpu_trojan::subsec::attacker_model}
Given that the target machine-learning model is usually not known during hardware trojan insertion, an attacker implanting a machine-learning backdoor through a trojanized hardware accelerator must always exploit at least two attack vectors.

First, they must be capable of implanting a programmable hardware trojan into an accelerator for machine learning before or during manufacturing. 
The hardware design process comprises multiple stages and involves a variety of stakeholders situated across the globe, opening up a multitude of attack vectors.
Before manufacturing, design files are sent between companies and third-party \ac{IP} cores, \ie, design files of self-contained hardware components crafted by dedicated \ac{IP} vendors, are used to speed up the development of larger \acp{SoC}.
For example, a machine-learning accelerator may be designed as a third-party \ac{IP} core and shipped to the integrator.
The final design comprises the individual components and is subsequently synthesized to a gate-level circuit description.
As hardware manufacturing is often outsourced, this circuit description is sent to a \textit{fab} that finally produces the \ac{IC}.
Hence, a supply chain attack could be conducted by the designer themselves by manipulating design files, the third-party \ac{IP} vendor by supplying a trojanized \ac{IP} core, an independent entity intercepting design files during transmission, or the \ac{IC} manufacturer by inserting manipulations before fabrication, all of which are common threat models in hardware trojan research~\cite{DBLP:conf/ches/SkorobogatovW12,DBLP:conf/sp/SturtonHWK11,DBLP:conf/ches/BeckerRPB13,DBLP:conf/sp/PuschnerMBKMP23}.
A single rogue entity often suffices for a successful trojan attack.

Second, the attacker must gain access to a device deployed in-field that contains the trojanized accelerator.
They must then extract the learning model~\cite{DBLP:conf/uss/SunSLM21}, insert a minimal backdoor, and program the backdoor to the trojanized accelerator, thereby activating the trojan.
In the case of a car, this could be done during a routine inspection, by breaking into the car, gaining remote access, or infiltrating the deployer of the learning model.
An attacker might want to provision a hardware trojan in \emph{all} vehicles, but upload the fatal backdoor only to selected targets.
For a successful attack, the adversary does \textit{not} require any knowledge of the training data.

Our attacker model implies significant capabilities. 
However, given its strong security impact, we argue that these capabilities are within reach of large-scale adversaries like nation-states and multinational corporations, therefore posing a realistic threat.
This especially becomes apparent when considering military~\cite{asaro_2012} and aerospace~\cite{DBLP:conf/wmcsa/KothariLL20} applications, in which machine-learning and hardware acceleration thereof are increasingly utilized for mission-critical functionalities.
Please note that manipulation of the hardware and the backdoor construction can be conducted by different entities with no detailed knowledge of the other attack stages.

\begin{figure*}
    \centering
    \includegraphics[width=0.7\textwidth]{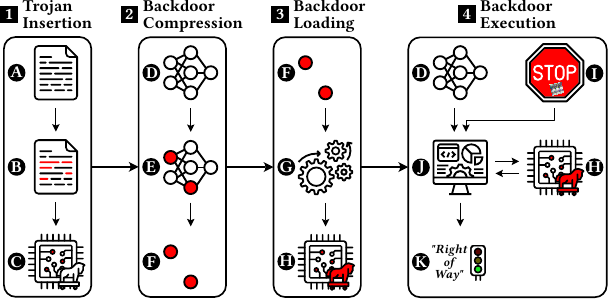}
    \caption{The four stages of our proposed trojan attack.}
    \label{dpu_trojan::fig::attack_flow}
\end{figure*}

\subsection{Attack Outline}
\label{dpu_trojan::subsec::attack_overview}

\autoref{dpu_trojan::fig::attack_flow} shows a detailed overview on the processing steps of our attack along the four stages outlined in \autoref{dpu_trojan::fig::attack_flow_overview}.

\par\smallskip\textbf{\circledtext[boxtype=O]{1} Trojan Insertion.}
Having access to the design files or circuit descriptions of the machine-learning accelerator~\circledtext{A}, the attacker inserts a programmable trojan~\circledtext{B}.
This trojan is designed to swap specific parameters while they are streamed to the accelerator to insert a minimal backdoor.
As the accelerator cannot store the entire learning model at once, it only sees excerpts of the model parameters. 
Also, it has no understanding of the model architecture.
Hence, the trojan needs to decide for itself when to replace the incoming parameters, without knowing their context.
To minimize the attack footprint, only very few parameters shall be replaced.
At this stage, the attacker only adds circuitry to store, locate, and exchange affected parameters, but does not yet load the manipulated parameters.
The \trojan thus remains inactive until it is programmed with the backdoor.
For this, the attacker provisions a programming interface that enables loading the manipulated parameters to the hardware even after deployment.
Finally, the trojanized accelerator is manufactured~\circledtext{C} by following the regular hardware design and manufacturing process.

\par\smallskip\textbf{\circledtext[boxtype=O]{2} Backdoor Compression.} 
The attacker gains access to the trained (and potentially quantized) learning model~\circledtext{D} of a traffic sign recognition system.
Using a copy of the original model, they implant a backdoor mechanism resulting in a backdoored learning model~\circledtext{E}.
Whenever a specific \emph{trigger} pattern is present in the input image of a source class (\eg, \enquote{stop sign}), the backdoored model will predict a specific target class (\eg, \enquote{right of way}) with high probability.
Since our hardware trojan mandates that only a minimal number of model parameters be altered, we propose a novel backdoor class that penalizes a large number of parameter changes.
Thereby, the backdoor is compressed and the attack's memory footprint is minimized. 
Finally, the attacker compares the original model and the backdoored one to extract the parameters~\circledtext{F} to be replaced by the trojan.

\par\smallskip\textbf{\circledtext[boxtype=O]{3} Backdoor Loading.} 
To arm the hardware \trojan, the attacker converts the modified parameters~\circledtext{F} to the format that is used by the hardware accelerator.
Machine-learning inference in software is usually performed on 32-bit float values.
However, as these are inefficient in hardware, quantization~\cite{DBLP:journals/corr/abs-2004-09602,DBLP:conf/cvpr/WangLLLH19,DBLP:conf/cvpr/JacobKCZTHAK18,DBLP:conf/aaai/ZhouMCF18} is often employed to reduce the bit width and instead operate on fixed-point values.
After making respective adjustments~\circledtext{G}, the attacker programs the corresponding values into the accelerator using the provisioned programming interface.
From now on, the trojan is active and will deploy the backdoor parameters whenever the target model is executed on the trojanized hardware accelerator~\circledtext{H}. 

\par\smallskip\textbf{\circledtext[boxtype=O]{4} Backdoor Execution.}
During inference, the original model~\circledtext{D} is executed in-field by a machine-learning software~\circledtext{J} on the victim system, \eg, an \ac{ECU} in a car, to perform classification tasks on input data~\circledtext{I} such as pictures of traffic signs.
To perform inference efficiently, the software makes use of the (trojanized) hardware accelerator~\circledtext{H} and streams to it the model parameters and input data over a sequence of computations. 
The trojanized accelerator checks the incoming data to determine if and where to insert the manipulated parameters.
If the data matches an entry in a list of manipulations, the \trojan substitutes the respective parameter before the requested computation is executed.
Once programmed, the trojan is only activated if the target model is streamed to the accelerator.
For every other learning model, it remains dormant.
As a result, the hardware (and thereby also the software) operates on a backdoored learning model and returns a malicious prediction~\circledtext{K}.
Input images without the trigger are correctly classified, while those that contain the trigger are falsely classified to the target class, namely \enquote{right-of-way}.
Note that the manipulation is performed entirely within the hardware---hidden from the victim who seemingly executes a trojan-free model.
Cryptographic checks applied to the model are ineffective in detecting our attack, as the model remains unaltered outside the accelerator.
\section{Minimal Backdoors}\label{dpu_trojan::sec::ml_trojan}

For a successful hardware trojan attack, the attacker must specify the model parameters to be manipulated as well as their new (malicious) values. 
Our trojan requires the backdoor to be realized by exchanging as few parameters as possible while still ensuring a reliable backdoor. 
Hence, we construct a \emph{minimal backdoor}, which builds on a regularized and sparse update of model parameters.

\subsection{From Learning to Backdoors}
Before presenting minimal backdoors, we briefly describe the learning process of neural networks and how it can be adapted to include backdoor functionality.

\par\smallskip\textbf{Neural Networks.}
A neural network for classification is a parameterized function $f_\theta(x)$ that processes an input vector $x \in \mathbb{R}^d$ and maps it to one of $c$ classes. 
The model parameters $\theta\in\mathbb{R}^m$ (or weights) define the network structure and control its computations. 
In supervised learning, they are determined based on training data $D=\big\{(x_i,y_i)\big\}_{i=1}^n$ consisting of $n$ examples $x_i$ with labels $y_i$. 
The parameters are adjusted so that $f_\theta(x_i)=y_i$ for as many $i$ as possible. 
This is achieved by optimizing a loss function $\ell\big(f_\theta(x) ,y,\theta\big)$ that measures the difference between a prediction $f_\theta(x)$ and the true label $y$. 
The optimal parameters $\theta^*$ can thus be defined as 
\[
\theta^*=\argmin_{\theta\in\mathbb{R}^m}L(\theta,D)=\argmin_{\theta\in\mathbb{R}^m}\sum_{i=1}^n \ell\big(f_\theta(x_i), y_i,\theta\big).
\]

For deep neural networks, solutions for $\theta^*$ can only be obtained approximately by
using training algorithms like \ac{SGD} which compute
\[
\theta_{t+1} = \theta_t - \tau \nabla_\theta \ell\big(f_\theta(x_j), y_j, \theta\big)
\]
for every pair $(x_j,y_j)$. That is, the parameters are adjusted by moving them into the direction of the steepest descent of $\ell$ weighted by the learning rate $\tau$ until the total loss $L$ converges.

\par\smallskip\textbf{Quantization.}
On hardware, the model $\theta$ is often \textit{not} provided in a standard format, such as 32-bit floating point numbers. Instead, the parameters are typically reduced in size and precision, a process called \emph{quantization}~\cite{DBLP:conf/cvpr/JacobKCZTHAK18,DBLP:journals/corr/abs-2004-09602}.
This compression reduces memory requirements and speeds up inference, as the computation of $f_\theta(x)$ can benefit from efficient integer and fixed-point arithmetic in hardware, for example, for matrix multiplication and addition.

Given a bit width $b$, quantization maps the model parameters from their original range $[\alpha, \beta]$ to integers in $[-2^{b-1}, 2^{b-1}-1]$. 
Let us denote the standard floor function by $\lfloor x\rfloor$, the scale as $s=(\beta-\alpha)/(2^{b}-1)$, and the zero point by $p_0=-\big\lfloor\alpha\cdot s\big\rfloor - 2^{b-1}$. 
An affine quantization of a real number $a$ is then defined as

$$q(a) = \Big\lfloor\frac{a}{s}+p_0\Big\rfloor_b$$
with the inverse mapping being 
$r(q) = \big(q-p_0) s$.
Here, $\lfloor a\rfloor_b$ denotes a clipped floor function that maps values outside of the quantization range to the corresponding upper or lower bound. 
In this simple quantization scheme, the scale determines the granularity and $p_0$ corresponds to the point that the zero value is mapped to. 
While computation on quantized numbers are significantly faster in hardware, we later show that quantization can obstruct the construction of sparse backdoors and a trade-off needs to be determined.  

\par\smallskip\textbf{Machine Learning Backdoors.}
Backdoors are a well-known security threat in machine learning.
The goal of these attacks is to make a learning model predict a selected class~$y_t$ whenever a given trigger $T$ is present in the input. 
If the attacker can manipulate the training data, they can easily insert examples of the form $(x+T,y_t)$ where the trigger $T$ is added to the
inputs~\cite{DBLP:journals/access/GuLDG19}. 
However, in our setting, only the model parameters can be modified and hence more recent backdooring techniques must be
applied~\cite{DBLP:conf/nips/ShafahiHNSSDG18,DBLP:conf/ccs/YaoLZZ19,DBLP:conf/ndss/LiuMALZW018}. In particular, our attack generates artificial input vectors
$\tilde{x}$ activating selected classes of the neural network and performs \ac{SGD} updates with $(\tilde{x}, y)$ and $(\tilde{x}+T, y_t)$ to create a
backdoored model~\cite{DBLP:journals/corr/SimonyanVZ13,ErhBenCou+09}.

\par\smallskip\textbf{Crafting Minimal Backdoors}
\label{sect:ml-backdoor-approach}
Finding a minimal backdoor can be phrased as an optimization problem aiming to determine a minimal parameter change $\delta$ that is added to the original parameters $\theta^*$, so that the backdoor becomes active in presence of the trigger $T$. 
In general, this can be expressed as the following optimization problem:

\begin{equation}
\begin{aligned}
&\min_{\delta}        &\qquad& \lVert\delta\rVert_0\\
&\text{s.t.} &      & f_{\theta^*+\delta}(x) = y_s, \\
&                  &      & f_{\theta^*+\delta}(x+T) = y_t \qquad \forall x\in F.
\end{aligned}
\label{eq:minimal-backdoor}
\end{equation}
Here, $F$ is a set of data points from the source class, $T$ is the trigger that is added to an image, $y_s$
is the source class and $y_t$ is the target class, which the
\trojan shall predict if the trigger is present, and $\lVert\delta\rVert_0$ is the number of entries in $\delta$ that are non-zero.
\autoref{eq:minimal-backdoor} is related to adversarial examples~\cite{DBLP:conf/sp/Carlini017,DBLP:journals/corr/GoodfellowSS14} but aims for a minimal perturbation to the \emph{model parameters} instead of the input $x$.

\par\smallskip\textbf{Backdoor Insertion.} 
To insert the backdoor, we fine-tune the parameters $\theta^*$ by using the samples in $F$ to obtain a solution for
\autoref{eq:minimal-backdoor} by solving
\begin{equation}
    \argmin_{\theta\in\mathbb{R}^m}\sum_{x\in F} \ell\big(\Tilde{f}_\theta(x), y_s,\theta\big) + \ell\big(\Tilde{f}_\theta(x+T), y_t,\theta\big),
\label{opt-problem-trojan}
\end{equation}
where $\Tilde{f}$ indicates that all layers except the final one are frozen. 

Similar to Liu~\etal~\cite{DBLP:conf/ndss/LiuMALZW018}, we design the trigger $T$ to boost the activation of a single neuron in the network.
This is advantageous when aiming for minimal backdoors for multiple reasons:
First, the highly excited neuron leads to sparser parameter changes since the majority of changes relate to this neuron. 
Second, freezing all but the final layer prevents many parameter changes that would otherwise be induced during optimization. 
To further minimize the backdoor, we use adaptive neuron selection, update regularization, and backdoor pruning, all of which we explain in the following.

\par\smallskip\textbf{Adaptive Neuron Selection.}
At the heart of the attack from Liu \etal~\cite{DBLP:conf/ndss/LiuMALZW018} is a neuron that is overexcited in presence of the trigger. 
They suggest to target the neuron with highest connectivity, \ie, if the weights $w_{1,i},\dots,w_{M,i}$ are connections to a neuron $n_i$ in the target layer, we choose $n_k$ with $k = \max_{i} \sum_{j} \lvert w_{j,i}\rvert.$ 
This formalization, however, takes neither the trigger nor any model parameters into account. 
Therefore, we propose an \emph{adaptive neuron selection} scheme leveraging gradient information to find an optimal neuron with respect to a given trigger and model.
To this end, we place the trigger $T$ on an empty image and compute 
$$a_j = \sum_i \Big\lvert \frac{\partial n_j}{\partial t_i}\Big\rvert$$
for every target neuron $n_j$, where $t_i$ are the pixels of $T$.
We choose the neuron with the highest $a_j$ over all $j$.
This corresponds to the neuron that can be best \emph{influenced} by the trigger and model at hand, thus requiring minimal changes to be adapted to our backdoor.

\par\smallskip\textbf{Update Regularization.} 
In order to change as few parameters as possible, we solve the modified optimization problem
\begin{equation}
    \begin{split}
        \argmin_{\delta\in\mathbb{R}^m}\sum_{x\in F} &\ell\big(\Tilde{f}_{\theta^*+\delta}(x), y_s,\theta^*\big)\\
        &+ \ell\big(\Tilde{f}_{\theta^*+\delta}(x+T), y_t,\theta^*\big) + \lambda \lVert \delta\rVert_p,
    \end{split}
    \label{opt-problem-trojan-reg-2}
\end{equation}
which penalizes deviations of the new model parameters from $\theta^*$. 
Natural choices for $p$ are $\{0,1,2\}$ where each $L^p$ norm leads to different behavior.
For $p\in\{1,2\}$, the regularization penalizes \emph{large} deviations from $\theta^*$ whereas $p=0$ allows unbounded deviations but penalizes \emph{every} existing deviation.
We will later see how the choice of $p$ affects the optimization.

\autoref{opt-problem-trojan-reg-2} can be optimized with \ac{SGD} for $p\in\{1,2\}$.
For $p=0$, however, the regularization term is not differentiable anymore.
Although removing neurons~\cite{DBLP:conf/nips/CunDS89,DBLP:conf/nips/LouizosUW17,DBLP:conf/nips/WenWWCL16} or weights~\cite{DBLP:journals/corr/HanMD15,DBLP:conf/iclr/UllrichMW17,DBLP:conf/icml/MolchanovAV17} of a network---also called \emph{pruning}---is connected to minimizing the $L^0$ norm, such approaches are often performed \emph{post training}.
Instead, for backdoor insertion, we perform $L^0$ regularization during optimization~\cite{DBLP:conf/iclr/LouizosWK18,DBLP:conf/cvpr/SrinivasSB17}.
We follow Louizos~\etal~\cite{DBLP:conf/iclr/LouizosWK18} and transform the parameters using \emph{gates} $z$ by computing the element-wise product $\Tilde{\theta}=z\odot\theta$. 
These gates are random variables with a density function parameterized by $\pi$.
The density is chosen such that $\pi$ can change the distribution to have most of its mass either at $1$ or $0$ to turn the gates \enquote{on} or \enquote{off}, respectively.
As long as the density is continuous, the value of $\pi$ for each parameter can be incorporated into the optimization problem to ensure that as few gates as possible are turned \enquote{on}.
After optimization, we sample the binary gates to obtain a final mask for the last layer.

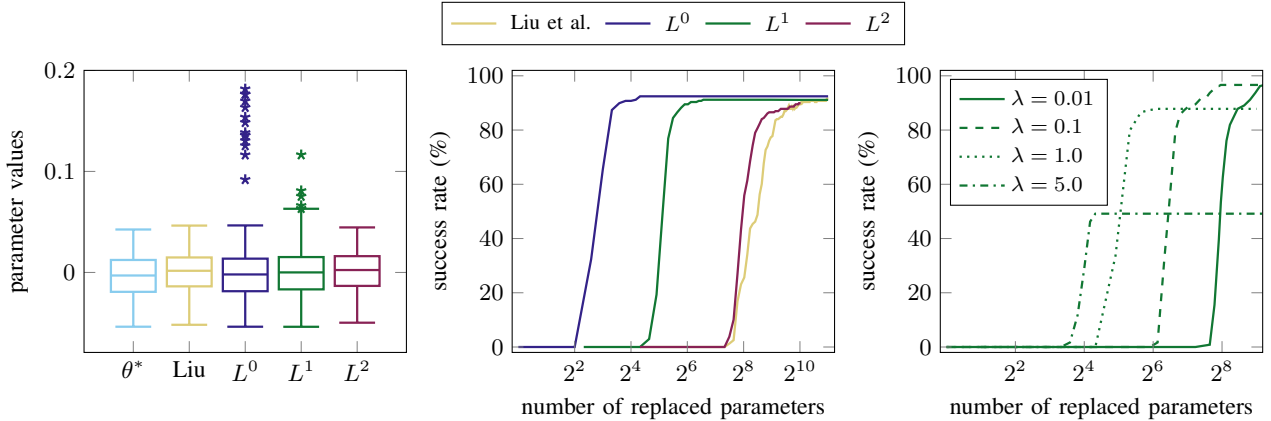
\begin{figure*}[ht]
    \centering  \small


\definecolor{col1}{HTML}{332288}
\definecolor{col2}{HTML}{117733}
\definecolor{col3}{HTML}{88CCEE}
\definecolor{col4}{HTML}{DDCC77}
\definecolor{col5}{HTML}{882255}

\begin{tikzpicture}

\pgfplotsset{
compat=1.11
}

\begin{groupplot}[group style = {group size = 3 by 1,
                                 horizontal sep = 40pt,
                                 vertical sep=20pt},
                                 tick label style={font=\small},
                                 width = 0.34\linewidth,
                                 height = 0.25\linewidth
                ]

    \nextgroupplot[
    boxplot/draw direction = y,
    xtick={1,2,3,4,5},
    xticklabels={$\theta^*$, Liu, $L^0$, $L^1$, $L^2$},
    ymax=0.2,
    ylabel=parameter values,
    ]
    \addplot+[boxplot,color=col3, line width=0.3mm] table[y=original, col sep=comma]{data/ml-trojan/weight-diff-results.csv};
    \addplot+[boxplot, color=col4, mark=star, mark options={col4}, line width=0.3mm] table[y=Liu, col sep=comma]{data/ml-trojan/weight-diff-results-liu.csv};
    \addplot+[boxplot,color=col1, mark=star, mark options={col1}, line width=0.3mm] table[y=L0, col sep=comma]{data/ml-trojan/weight-diff-results.csv};
    \addplot+[boxplot,color=col2, line width=0.3mm] table[y=L1, col sep=comma]{data/ml-trojan/weight-diff-results.csv};
    \addplot+[boxplot,color=col5, line width=0.3mm] table[y=L2, col sep=comma]{data/ml-trojan/weight-diff-results.csv};

    \nextgroupplot[
      title = {},
      xlabel={\# replaced parameters},
      ylabel={success rate (\%)},
      axis on top,
      grid style={on layer=axis background},
      label style={font=\small},
      xtick={4,16,64,256,1024, 4096},
      ymin=0, ymax=100,
      xmode=log,
      log basis x={2},
      tick label style={font=\small},
      legend style = { column sep = 3pt, legend columns = 4, at={(0.5, 1.35)}, anchor=north},
      enlarge x limits=0.02,
      enlarge y limits=0.02,
      grid=both
    ]

\addplot[color=col4, line width=0.3mm] plot [] table [x=n_replacements, y expr=\thisrow{success-rate}*100, col sep=comma]{data/ml-trojan/results_pruning_vgg16-1024_trojaned_300epochs_-SGD-LR-0.0001-lambda-0.01-p-None_final.csv};
\addlegendentry{Liu \etal}

\addplot[color=col1, line width=0.3mm] plot [] table [x=n_replacements, y expr=\thisrow{success-rate}*100, col sep=comma]{data/ml-trojan/results_pruning_vgg_model_1024_unitsL0-backdoor.csv};
\addlegendentry{$L^0$}

\addplot[color=col2, line width=0.3mm] plot [] table [x=n_replacements, y expr=\thisrow{success-rate}*100, col sep=comma]{data/ml-trojan/results_pruning_model_trojaned_300epochs_-SGD-LR-0.0001-lambda-0.5-p-1.csv};
\addlegendentry{$L^1$}

\addplot[color=col5, line width=0.3mm] plot [] table [x=n_replacements, y expr=\thisrow{success-rate}*100, col sep=comma]{data/ml-trojan/results_pruning_model_trojaned_300epochs_-SGD-LR-0.0001-lambda-3.0-p-2.csv};
\addlegendentry{$L^2$}

\nextgroupplot[
      title = {},
      xlabel={\# replaced parameters},
      ylabel={success rate (\%)},
      axis on top,
      grid style={on layer=axis background},
      label style={font=\small},
      xtick={4,16,64,256},
      tick label style={font=\small},
      xmax=512,
      ymin=0, ymax=100,
      xmode=log,
      log basis x={2},
      legend style = {font=\scriptsize, yshift=0.1cm},
      legend pos=north west,
      legend cell align={left},
      enlarge x limits=0.02,
      enlarge y limits=0.02,
      grid=both
    ]

\addplot[color=col2, line width=0.3mm] plot [] table [x=n_replacements, y expr=\thisrow{success-rate}*100, col sep=comma]{data/ml-trojan/results_pruning_model_trojaned_250epochs_-Adam-LR-0.0001-lambda-0.01-p-1-train_data-1.0.csv};
\addlegendentry{$\lambda=0.01$}

\addplot[color=col2, line width=0.3mm, dashed] plot [] table [x=n_replacements, y expr=\thisrow{success-rate}*100, col sep=comma]{data/ml-trojan/results_pruning_model_trojaned_250epochs_-Adam-LR-0.0001-lambda-0.1-p-1-train_data-1.0.csv};
\addlegendentry{$\lambda=0.1$}


\addplot[color=col2, line width=0.3mm, dotted] plot [] table [x=n_replacements, y expr=\thisrow{success-rate}*100, col sep=comma]{data/ml-trojan/results_pruning_model_trojaned_175epochs_-SGD-LR-0.0001-lambda-1.0-p-1.csv};
\addlegendentry{$\lambda=1.0$}

\addplot[color=col2, line width=0.3mm, dashdotted] plot [] table [x=n_replacements, y expr=\thisrow{success-rate}*100, col sep=comma]{data/ml-trojan/results_pruning_model_trojaned_250epochs_-Adam-LR-0.0001-lambda-5.0-p-1-train_data-1.0.csv};
\addlegendentry{$\lambda=5.0$}

\end{groupplot}
\end{tikzpicture}
    \caption{
        \textit{Left:} Box-plot of the parameter distribution in the final layer before and after backdoor insertion. 
        \textit{Mid:} Evolution of the backdoor success rate for different values of $p$ when replacing parameters of the original model from largest to smallest difference.
        \textit{Right:} Evolution of the backdoor success rate for $p=1$ and different regularization strengths $\lambda$.
    }
    \label{fig:eval-overview}
\end{figure*}

\par\smallskip\textbf{Backdoor Pruning.}
Solving the optimization problem in~\autoref{opt-problem-trojan-reg-2} yields a vector $\delta$ of parameter changes that can be added to the original parameters $\theta^*$ to obtain a backdoored model. 
However, not every parameter change in $\delta$ is required for an effective backdoor. 
To find the minimal number of required parameter changes, we \emph{prune} the parameters of the backdoored model:
First, we sort the parameter changes $\lvert\delta\rvert$ in decreasing order to obtain $\delta^{(1)},\dots,\delta^{(m)}$. 
Starting with $\delta^{(1)}$, we sequentially add changes to the corresponding parameters in $\theta^*$ to obtain a new model between $\theta^*$ and $\theta^*+\delta$. 
Using unseen data, we compute the \emph{success rate} (\ie, the fraction of data which is classified as $y_t$ when the trigger is present) and the \emph{accuracy}. 
Thereby, we determine the optimal number of parameter changes as the backdoor effectiveness increases continuously.

\subsection{Evaluation}
\label{dpu_trojan::subsec::ml_eval}
Once the backdoor is inserted, it remains to evaluate the manipulated model against two criteria.
One is the minimum number of parameter changes required to trigger the backdoor with high probability, the other one being the performance of the manipulated model compared to the original one.

\par\smallskip\textbf{Dataset and Models.}
We use the German Traffic Sign dataset~\cite{DBLP:conf/ijcnn/HoubenSSSI13} to simulate our attack in an automtotive setting.
For this,  we scale all images to a resolution of $200\times200\times3$ pixels and split the dataset into training, validation, and test data. 
For now, the trigger size is fixed to $30\times30\times3$ pixels (2.25\% of the image area) and we train a VGG16 model~\cite{DBLP:journals/corr/SimonyanZ14a} with $1024$ dense units in the final layers. 

Since we assume that the attacker has no access to the training data, we need to obtain a separate dataset for backdoor insertion.
While Liu~\etal~\cite{DBLP:conf/ndss/LiuMALZW018} create artificial training images, we take $30$ additional pictures of stop signs in our local city and insert the backdoor by solving the optimization problem in \autoref{opt-problem-trojan-reg-2} using \ac{SGD} optimization for $300$ epochs.
We select \ac{SGD} optimization, because other optimization algorithms like RMSProp or Adam produced significantly more parameter changes in our experiments. 
We also find that the regularization strength $\lambda$ and learning rate $\tau$ are hyperparameters that influence the sparsity of the backdoor
and hence have to be calibrated.
For this, we perform a grid search in $[0.01, 5]$ for $\lambda$ and $[0.0001, 0.001]$ for  $\tau$.

\par\smallskip\textbf{Parameter Distribution Change.}
When inspecting the changes to the clean model $\theta^*$ induced by the backdoor, we find that the majority of them affect parameters connected to the output neuron of class $y_t$.
This is not true for the baseline approach of Liu~\etal~\cite{DBLP:conf/ndss/LiuMALZW018}, which induces larger changes to other parameters as well. 
Figure~\ref{fig:eval-overview} (left) depicts a boxplot of the parameter distribution of the target layer that has been chosen for backdoor insertion for $\theta^*$ and the backdoored models in respect to different regularization norms. 
For $p\in\{0,1\}$, we observe parameter outliers compared to the distribution of 
$\theta^*$, \ie, the optimization induces larger weight changes to insert the backdoor. 
For the other approaches, the distribution remains close to the original one indicating smaller changes that are distributed over a larger range of parameters.

\par\smallskip\textbf{Sparsity.}
\autoref{fig:eval-overview} (mid) shows the evolution of the trigger success rate when following our pruning approach.
This confirms observations from the parameter distributions in the pruning process: 
$L^0$ and $L^1$ regularization induce larger parameter changes on fewer parameters and achieve sparser backdoors. 
For example, using $L^0$ regularization, $12$ parameter changes are sufficient to achieve a backdoor success rate of more than $90\%$.
The approach of Liu~\etal~\cite{DBLP:conf/ndss/LiuMALZW018} induces more
than $1000$ weight changes and thereby exhibits the highest change ratio of all methods.
Furthermore, the final success rate of the regularized backdoor does not reach 100\%. 
As shown in \autoref{fig:eval-overview} (right) for $p=1$, it is bounded by the regularization strength $\lambda$. 
Hence, the attacker must balance the trade-off between backdoor sparsity and success rate. 
To facilitate comparability, we propose a \ac{DSR} of $90\%$ and measure the \emph{sparsity} $\Delta\mathcal{S}$ of the backdoors as the minimum number of parameter changes required to obtain the \ac{DSR}. 

\par\smallskip\textbf{Quantization as a Hurdle.}
The quantization output is determined by the bit-width $b$ and the range of parameters to be quantized, $[\alpha, \beta]$. 
These parameters determine the discrete $2^b-1$ \emph{bins} between $\alpha$ and $\beta$ into which the floating-point values are assigned.
From the parameter distribution in \autoref{fig:eval-overview}, we see that quantization can be obstructive for our attack because a large parameter change, as observed for $L^0$ regularization, can significantly affect $\beta$ and thereby the entire quantization output. 
Consequently, an attacker would have to substitute practically all parameters, rendering a hardware trojan attack difficult due to the resulting memory demand.
We denote by $\Delta\mathcal{Q}$ the total number of parameters that are changed after performing quantization on the model containing the backdoor. 
Ideally, we have $\Delta\mathcal{S} = \Delta\mathcal{Q}$, \ie, the quantization of the model does not further impact the sparsity of the backdoor.
If $\Delta\mathcal{S} < \Delta\mathcal{Q}$, quantization increases the number of parameter changes, thereby reducing the stealthiness and memory efficiency of the attack. 
To compute $\Delta\mathcal{Q}$, we use the quantizer shipped with the Xilinx Vitis AI toolkit in its standard configuration and count the differences in bytes.

\begin{table*}[ht]
    \caption{
        Impact of \subref{dpu_trojan::subtab::trigger_size_eval}~trigger size and \subref{dpu_trojan::subtab::model_archtecture_eval}~model type on the difference in test accuracy $\Delta\mathcal{A}$ in percentage points, sparsity $\Delta\mathcal{S}$ for a \ac{DSR} of 90\%, and parameter changes after quantization $\Delta\mathcal{Q}$ using different regularization techniques. 
    }
    \label{dpu_trojan::tab::regularization_eval}
    \begin{subtable}[c]{\textwidth}
    \centering
        \begin{tabularx}{\textwidth}{Y|C{8mm}C{8mm}C{1cm}|C{8mm}C{8mm}C{1cm}|C{8mm}C{8mm}C{1cm}|C{8mm}C{8mm}C{1cm}}
            \toprule
            \multirow[t]{2}{*}{\textbf{Trigger Size}} & \multicolumn{3}{c|}{\textbf{Liu~\etal}} & \multicolumn{3}{c|}{\textbf{$L^0$ Regularization}} &\multicolumn{3}{c|}{\textbf{$L^1$ Regularization}} & \multicolumn{3}{c}{\textbf{$L^2$ Regularization}}\\
            \cmidrule{2-13}
            (\% of image)& $\Delta\mathcal{A}$& $\Delta\mathcal{S}$ & \textbf{$\Delta\mathcal{Q}$} & $\Delta\mathcal{A}$ & $\Delta\mathcal{S}$ & $\Delta\mathcal{Q}$& $\Delta\mathcal{A}$ & $\Delta\mathcal{S}$ & $\Delta\mathcal{Q}$& $\Delta\mathcal{A}$ & $\Delta\mathcal{S}$ & $\Delta\mathcal{Q}$\\
            \midrule
            $20\times 20\:(1.00\%)$ & 1.84\% & 1339 & 1339 & 1.15\% & 139 & \numprint{43739} & 0.21\% & 617 & 617 & 0.18\% & 813 & 813\\
            $30\times 30\:(2.25\%)$ & 1.48\% & 1092 & 1092 & 0.09\% & 13 & \numprint{43739} & 0.05\% & 80 & 80 & 0.08\% & 202 & 202\\
            $40\times 40\:(4.00\%)$ & 0.05\% & 87 & 87 & 0.20\% & 3 & \numprint{43739} & 0.02\% & 63 & 63 & 0.00\% & 74 & 74\\
            $50\times 50\:(6.25\%)$ & 0.11\% & 60 & 60 & 0.48\% & 2 & \numprint{43739} & 0.00\% & 7 & 7 & 0.00\% & 12 & 12\\
            \bottomrule
        \end{tabularx}
        \subcaption{Impact of the trigger size on the backdoor properties for a VGG-16 network.}
        \label{dpu_trojan::subtab::trigger_size_eval}
    \end{subtable}
    \\

    \begin{subtable}{\textwidth}
    \centering
        \begin{tabularx}{\textwidth}{Y|C{8mm}C{8mm}C{1cm}|C{8mm}C{8mm}C{1cm}|C{8mm}C{8mm}C{1cm}|C{8mm}C{8mm}C{1cm}}
            \toprule
            \multirow[t]{2}{*}{\textbf{Model}} & \multicolumn{3}{c|}{\textbf{Liu~\etal}} & \multicolumn{3}{c|}{\textbf{$L^0$ Regularization}} &\multicolumn{3}{c|}{\textbf{$L^1$ Regularization}} & \multicolumn{3}{c}{\textbf{$L^2$ Regularization}}\\
            \cmidrule{2-13}
            & $\Delta\mathcal{A}$& $\Delta\mathcal{S}$ & \textbf{$\Delta\mathcal{Q}$} & $\Delta\mathcal{A}$ & $\Delta\mathcal{S}$ & $\Delta\mathcal{Q}$& $\Delta\mathcal{A}$ & $\Delta\mathcal{S}$ & $\Delta\mathcal{Q}$& $\Delta\mathcal{A}$ & $\Delta\mathcal{S}$ & $\Delta\mathcal{Q}$\\
            \midrule
            AlexNet & 0.20\% & 860 & 860 & 0.39\% & 19 & \numprint{174093} & 0.18\% & 654 & 654 & 0.05\% & 713 & 713\\
            VGG-13 & 1.44\% & 2018 & 2018 & 0.98\% & 7 & \numprint{173684} & 1.20\% & 564 & 564 & 1.20\% & 758 & 758\\
            VGG-19 & 1.46\% & 1366 & 1366 & 1.81\% & 10 & \numprint{176118} & 1.85\% & 499 & 499 & 1.38\% & 905 & 905\\
            \bottomrule
        \end{tabularx}
        \subcaption{Impact of different architectures on the backdoor for a fixed trigger size of $30\times30$ pixels.}
        \label{dpu_trojan::subtab::model_archtecture_eval}
    \end{subtable}
\end{table*}

\par\smallskip\textbf{Influence of Trigger Size, Model, and Dataset.}
In the following, we evaluate the influence of the trigger size, model architecture, and dataset on the sparsity $\Delta\mathcal{S}$, the number of parameter changes after quantization $\Delta\mathcal{Q}$, and the difference in test accuracy $\Delta\mathcal{A}$ compared to the original $\theta^*$, see \autoref{dpu_trojan::tab::regularization_eval} and \autoref{tab:face-recognition}.

\par\smallskip\textbf{Size of the Trigger.}
To measure the impact of the trigger size, we use triggers covering between $1\%$ and $6.25\%$ of the input images. The corresponding results are shown in \autoref{dpu_trojan::subtab::trigger_size_eval}.
Larger triggers ease hardware \trojan implementation, because sparsity and accuracy improve with increasing size of $T$. 
This confirms our observation that the target neuron can be excited stronger by larger triggers in the input. 
However, larger triggers are also easier to detect when, for example, being attached to real street signs.

$L^0$ regularization results in extremely sparse backdoors.
For example, only three changes are sufficient to achieve 90\% \ac{DSR} for a trigger covering $4\%$ of the input image.
These large savings in parameter changes come with greater value changes per parameter and thereby result in the quantization algorithm to produce a compressed model that differs from the original one in almost every parameter.
Hence, $L^1$ and $L^2$ regularization are a better fit since they reduce the parameter changes compared to the baseline method of Liu~\etal~\cite{DBLP:conf/ndss/LiuMALZW018} significantly while keeping value changes small enough to not impact quantization of unchanged parameters.

\par\smallskip\textbf{Model Architecture.}
Next, we investigate the influence of different model architectures, namely VGG-13~\cite{DBLP:journals/corr/SimonyanZ14a}, VGG-19~\cite{DBLP:journals/corr/SimonyanZ14a}, and AlexNet~\cite{DBLP:conf/nips/KrizhevskySH12}, for a trigger size of $30\times30$ pixels.
All three networks feature a different number of layers and $4096$ units in the final layer.
Hence, the number of potential target neurons is much larger compared to VGG-16.
From \autoref{dpu_trojan::subtab::model_archtecture_eval}, we observe that the generated backdoors are less sparse, likely due to the
higher number of neurons in the final layers. 
Using $L^1$ regularization saves between 24\% and 76\% parameter changes compared to Liu~\etal~\cite{DBLP:conf/ndss/LiuMALZW018} while
being resistant to quantization.
Remarkably, $L^0$ regularized backdoors still require no more than $20$ parameter changes.


\par\smallskip\textbf{Dataset.}
Finally, we apply our attack to a face recognition model by Parkhi~\etal~\cite{DBLP:conf/bmvc/ParkhiVZ15}, which was trained on $2.6$ million images. 
As this model comes with \numprint{2622} output classes, it has about $60\times$ more parameters in the final layer than the traffic sign models. 
Here, we create artificial images that are assigned to our source class with high probability~\cite{ErhBenCou+09} to conduct the fine-tuning from~\autoref{opt-problem-trojan-reg-2}.
We follow the work of Liu~\etal~\cite{DBLP:conf/ndss/LiuMALZW018} and use a trigger size of $60\times60$ pixels
($7\%$ of the input size) and report the results in \autoref{tab:face-recognition}.
\begin{table}[t]
    \small
    \centering
    \caption{Difference in test accuracy $\Delta\mathcal{A}$, sparsity $\Delta\mathcal{S}$, and changes after quantization $\Delta\mathcal{Q}$ for a face recognition dataset.}
    \begin{tabularx}{\linewidth}{X|C{0.15\linewidth}|C{0.15\linewidth}|C{0.2\linewidth}}
        \toprule
            &$\Delta\mathcal{A}$& $\Delta\mathcal{S}$ & \textbf{$\Delta\mathcal{Q}$}\\
            \cmidrule{1-4}
            Liu et al. & 0.12\%\ & 180 & 180\\
            $L^0$ Regularization & 4.01\% & 4 & \numprint{10606853}\\
            $L^1$ Regularization & 0.80\% & 5 & 5\\
            $L^2$ Regularization & 0.16\% & 341 & 341\\
        \bottomrule
    \end{tabularx}
    \label{tab:face-recognition}
\end{table}
Despite the optimization problem covering more than $10$ million parameters, the regularized backdoors are extremely sparse with only $5$ affected parameters for $L^1$ regularization,
even in presence of quantization. 
Compared to the baseline of Liu~\etal~\cite{DBLP:conf/ndss/LiuMALZW018}, the backdoor is compressed by $97$\%. 
We conclude that sparse backdoors exist independent of the dataset and model size.

\par\smallskip\textbf{Robustness to Parameter Changes.}
Our attacker model assumes that the adversary can deploy a backdoor for a specific learning model that is later executed on a trojanized machine-learning accelerator.
However, since the deployed model may change over time, \eg, because of fine-tuning as part of a software update, we investigate the implications of small parameter changes on the effectiveness of our backdoor.
To this end, we fine-tune the original model for $20$ epochs using \ac{SGD} and insert a backdoor after each epoch to evaluate our attack.
For fine-tuning, we utilize $70\%$ of our test data (\numprint{4400} images) and choose a learning rate that inflicts changes to the model but maintains its performance.

\begin{figure}[h!]
    \centering
    \definecolor{col1}{HTML}{332288}
\definecolor{col2}{HTML}{117733}
\definecolor{col3}{HTML}{88CCEE}
\definecolor{col4}{HTML}{DDCC77}
\definecolor{col5}{HTML}{882255}

\begin{tikzpicture}

  \centering
  \begin{axis}[
        ybar=4pt, axis on top,
        height=3.5cm, width=0.5\textwidth,
        bar width=0.25cm,
        enlarge x limits=0.25,
        ymin=50, ymax=100,
        log basis y={10},
        axis x line*=bottom,
        nodes near coords,
        nodes near coords align={vertical},
        point meta=rawy,
        tick label style={font=\footnotesize},
        label style={font=\footnotesize},
        legend style={
            at={(0.45,-0.5)},
            anchor=north,
            legend columns=-1,
            font=\footnotesize,
            /tikz/every even column/.append style={column sep=0.3cm}
        },
        legend entries={SR (Liu et al.), SR ($L^0$), SR ($L^1$), SR ($L^2$)},
        ylabel={Percentage (\%)},
        symbolic x coords={
           AlexNet,
           VGG-13,
           VGG-19},
       xtick=data,
       tick pos=left,
       nodes near coords style={font=\footnotesize},
       legend image code/.code={
        \draw [#1] (0cm,-0.1cm) rectangle (0.2cm,0.25cm); },
    ]
    \addplot [draw=none, fill=col4] coordinates {
      (AlexNet, 92)
      (VGG-13, 88) 
      (VGG-19, 90)
      };
    \addplot [draw=none, fill=col1] coordinates {
      (AlexNet, 91)
      (VGG-13, 89) 
      (VGG-19, 90)
      };
   \addplot [draw=none,fill=col2] coordinates {
      (AlexNet, 79)
      (VGG-13, 89) 
      (VGG-19, 90)
      };
   \addplot [draw=none, fill=col5] coordinates {
      (AlexNet, 90)
      (VGG-13, 88) 
      (VGG-19, 89)
      };
      \addlegendimage{line legend,black,dashed}

    \coordinate (A) at (axis cs:AlexNet,90);
    \coordinate (O1) at (rel axis cs:0,0);
    \coordinate (O2) at (rel axis cs:1,0);


  \end{axis}
  \end{tikzpicture}
    \caption{Mean success rate (SR) of the backdoor after fine-tuning for $20$ epochs.}
    \label{dpu_trojan:fig:stability-eval}
\end{figure}

\autoref{dpu_trojan:fig:stability-eval} depicts the mean success rate after fine-tuning three different learning models on unseen data for 20
epochs. We observe that the backdoors still maintain a high success rate despite the changes inflicted upon the model. 
Hence, our attack appears to be robust against parameter changes that could  occur in practice. 
Thus far, our trojan only becomes active if the original model is executed.
Given these results on the backdoor robustness, this rule could be relaxed so that the trojan activates even if only the parameters' most significant bits match those of the original model.
\section{Case Study with the Xilinx Vitis AI}
\label{dpu_trojan::sec::case_study}
We demonstrate our attack using Xilinx Vitis~AI~\cite{vitis-ai} for inference acceleration on a Zynq UltraScale+ MPSoC ZCU104 device.
We chose this \ac{FPGA} platform for demonstration as it can be employed for safety-sensitive applications and, at the same time, is accessible to researchers.
Also, importantly, our \ac{FPGA} case study is a good approximation of a similar \ac{ASIC}-based trojan, which could be employed in high-volume applications. 
Google's TPU~\cite{DBLP:conf/isca/JouppiYPPABBBBB17}, for example, inhibits an architecture similar to the one of Xilinx.

\subsection{\acs{DPU} Architecture}
\label{dpu_trojan::subsec::dpu_architecture}
Xilinx Zynq UltraScale+ MPSoC devices combine a \acl{PS} based on ARM Cortex \acsp{CPU} with an \ac{FPGA}-typical \acl{PL} region.
External memory is part of the \acl{PS} but shared with the \acl{PL} via data and address buses.
The \acsp{CPU} are together referred to as \ac{APU}.
The \ac{DPU} is a commercial machine-learning accelerator \ac{IP} core that can be implemented in the \acl{PL}.
Its Verilog description is available on GitHub~\cite{vitis-ai-github} but is encrypted according to IEEE standard 1735~\cite{Automation2015}. 
However, this standard is susceptible to oracle attacks~\cite{DBLP:conf/ccs/ChhotarayNSFT17} and key extraction~\cite{DBLP:conf/sp/SpeithSEF0P22}. 
Hence, recovery, reverse engineering, manipulation, and subsequent re-encryption of the protected \ac{IP} is feasible.

\par
\begin{figure}[htb]
    \centering
    \includegraphics[width=0.8\linewidth]{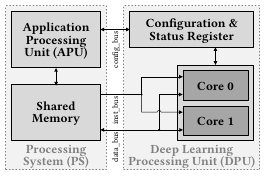}
    \caption{Top-level view of a \ac{DPU} with two processing cores and its connectivity to the \acl{PS}.}
    \label{dpu_trojan::fig::dpu_top}
\end{figure}

\smallskip\textbf{\acs{DPU}.} 
The \ac{DPU} accelerates inference computations such as convolutions and pooling.
To achieve this, it processes instructions to load, store, or operate on data.
The \ac{APU} controls the inference flow while off-loading computation-heavy tasks to the \ac{DPU}, which receives partial model parameters and inputs for the current layer but is unaware of their context.
The \ac{DPU} comprises one or more acceleration cores as well as shared configuration and status registers, see~\autoref{dpu_trojan::fig::dpu_top}.
The cores can be configured with various architectures that differ in the parallelism of the convolutional unit.
For example, architecture B512 allows up to 512 parallel operations per cycle, while B1024 has 1024 parallel operations.
Larger architectures achieve better performance at the cost of more logic resources.
The \ac{DPU} communicates with the \acl{PS} via buses for configuration (\texttt{conf\_bus}), instructions (\texttt{inst\_bus}), and data (\texttt{data\_bus}).
Each core features one bus for instructions and one or more data buses.
In our case study, we employ the largest available architecture (B4096) in a single core \ac{DPU} configuration.

\par
\begin{figure}[htb]
    \centering
    \includegraphics[width=0.8\linewidth]{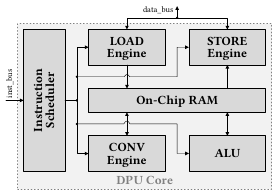}
    \caption{Inside view of a \ac{DPU} core with a single data port.}
    \label{dpu_trojan::fig::dpu_core}
\end{figure}

\smallskip\textbf{\acs{DPU} Core.} 
Within each \ac{DPU} core, the \texttt{inst\_bus} is connected to an instruction scheduler that controls the memory management and compute engines, see~\autoref{dpu_trojan::fig::dpu_core}.
The model parameters and input data for the current layer are transmitted from shared memory through the \texttt{data\_bus} that is connected to the LOAD and STORE engines.
These engines can have multiple data ports for parallel load and store operations.
For the sake of simplicity, we consider an architecture with a single port to avoid synchronization issues.

The data arriving through the LOAD engine is buffered in the on-chip \ac{RAM} for processing.
This makes the LOAD engine a promising attack target, as the buffer enables us to replace model parameters for backdoor insertion before the actual computation begins.
Once data has been written to the buffer, depending on the requested \ac{DPU} operation, either the CONV engine or the \ac{ALU} takes over.
The CONV engine is optimized for convolution and fully-connected layers, while the \ac{ALU} takes care of pooling and element-wise operations.
Once all computations on the buffered data are completed, the \ac{APU} instructs the STORE engine to write the results to shared memory.
During inference, the \ac{APU} iteratively queries the \ac{DPU} until all layers of the learning model have been processed.

\par
\begin{figure}[htb]
    \centering
    \includegraphics[width=0.9\linewidth]{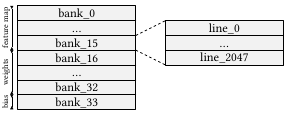}
    \caption{Logical memory layout of the on-chip \ac{RAM} for the employed \ac{DPU} configuration.}
    \label{dpu_trojan::fig::dpu_ram_layout}
\end{figure}

\smallskip\textbf{Logical Memory Layout.}
The \ac{DPU} on-chip memory is organized in \ac{RAM} banks comprising 2048 memory lines each, see~\autoref{dpu_trojan::fig::dpu_ram_layout}.
The number of banks and the size of each memory line depend on the \ac{DPU} architecture.
For B4096, there are 34 banks and each memory line is 16 bytes wide.
A bank is uniquely identified by the \texttt{bank\_id} and a memory line by the \texttt{bank\_addr}.
Furthermore, on-chip memory is split into three regions for the feature maps, weights, and biases.
The assignment of banks to regions is fixed.
For the target \ac{DPU} configuration, the first 16 banks are reserved for feature maps, the next 17 for weights, and the last one for biases.

\par
\begin{figure}[htb]
    \centering
    \includegraphics[width=0.9\linewidth]{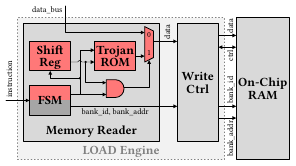}
    \caption{Simplified illustration of the \ac{DPU} LOAD engine including the added \trojan logic (in red).}
    \label{dpu_trojan::fig::dpu_load_engine}
\end{figure}

\smallskip\textbf{LOAD Engine.}
The LOAD engine retrieves data from shared memory, see \autoref{dpu_trojan::fig::dpu_load_engine} for a high-level overview.
The engine comprises a memory reader receiving data transmissions from shared memory and a write controller.
The memory reader \ac{FSM} parses load instructions received via the \texttt{inst\_bus} and passes \texttt{bank\_id}, \texttt{bank\_addr}, and the data from the \texttt{data\_bus} to the write controller.
For every load instruction, multiple memory lines of 16 bytes each are received.
The write controller forwards the signals to the on-chip \ac{RAM}, thereby writing the incoming data to this buffer.

\par
\begin{figure}[htb]
    \centering
    \includegraphics[width=0.9\linewidth]{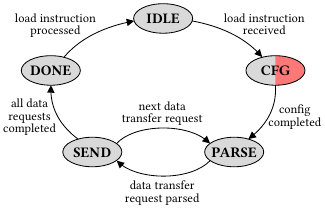}
    \caption{State graph of the \ac{FSM} controlling the memory reader of the LOAD engine in a \ac{DPU} core. Hardware \trojan logic is added to the \texttt{CFG} state.}
    \label{dpu_trojan::fig::ddr_reader_fsm}
\end{figure}

\smallskip\textbf{Memory Reader \acs{FSM}.}
The abstracted memory reader \ac{FSM} of the LOAD engine comprises five distinct states, see~\autoref{dpu_trojan::fig::ddr_reader_fsm}.
Some sub-states are omitted for clarity.
Once a new load instruction is received via the \texttt{inst\_bus}, the memory reader assumes the \texttt{CFG} state to receive data transmissions through the \texttt{data\_bus} in consecutive data transfers.
Among other information, a load instruction contains an address identifying the data source in shared memory (\texttt{ddr\_addr}) and the destination in the on-chip \ac{RAM} (\texttt{bank\_id} and \texttt{bank\_addr}).
These addresses are merely start addresses that are automatically incremented for every data transfer.
Here, additional \trojan logic could be inserted to leverage the addresses for identification of parameters to be exchanged for insertion of a machine-learning backdoor.
Once configuration in the \texttt{CFG} state is completed, the memory reader repetitively requests and parses data transfers in the \texttt{PARSE} and \texttt{SEND} states. 
Finally, the memory reader transitions to the \texttt{DONE} and subsequently the \texttt{IDLE} state and can then handle the next load instruction.

\subsection{Trojanizing the \acs{DPU}}
\label{dpu_trojan::subsec::trojanizing_dpu}

\par\smallskip\textbf{Trojan Insertion.}
Our programmable trojan resides in the memory reader of the LOAD engine, see \autoref{dpu_trojan::fig::dpu_load_engine}.
It comprises a \ac{ROM}, additions to an \ac{FSM}, a shift register, and a \ac{MUX}.
Some control logic is omitted for clarity.

Later on, the trojan \ac{ROM} will hold the manipulated parameters that realize the machine-learning backdoor.
Given the programmable nature of \acp{FPGA}, the \ac{ROM} can be updated via the bitstream.
Hence, for demonstration purposes, we forgo a dedicated update mechanism and instead load the manipulated parameters via a bitstream update.
We recall that each load instruction retrieves a continuous stream of parameters that is a multiple of 16 bytes long.
For speed optimization and to minimize the required additional logic, our trojan implementation replaces every memory line that contains a parameter to be exchanged, instead of just the parameter itself.
Because our backdoor requires only few parameter changes that often even reside within the data loaded by the same load instruction, the resulting memory overhead is negligible.

In addition to the manipulated parameters, the trojan stores shared memory addresses (\texttt{ddr\_addr}) used to identify the target load instructions.
Within the \texttt{CFG} state of the memory reader \ac{FSM}, we check the current \texttt{ddr\_addr} (from which data is about to be received) against the target addresses.
In case of a match, the trojan initiates exchanging incoming parameters with manipulated ones stored in the \ac{ROM}.
As these addresses are independent of the trojan logic, they can be updated similar to the \ac{ROM} contents.

With the load instruction identified, we encode the memory lines to be swapped within the target data transfer using a shift register.
Due to the limited number of parameter changes, not all of the 64 memory lines retrieved by one load instruction must be replaced. 
The shift register contains a \texttt{1} for each memory line to be exchanged and a \texttt{0} for every other line.
It is shifted for each data transfer, \ie, every received memory line.
The shift register output is used together with the \ac{FSM} output to activate the parameter exchange by controlling the \ac{ROM} and the \ac{MUX}.

Upon activation of the parameter exchange, the \ac{MUX} forwards the backdoor parameters obtained from the trojan \ac{ROM} to the write controller and finally to the on-chip \ac{RAM}.
Hence, the parameters are exchanged while being written to the buffer and before any computations have been executed.
Subsequent computations are thus performed on the manipulated parameters, \ie, \textit{using the backdoored learning model}.
These changes are invisible outside the accelerator.

\par\smallskip\textbf{Backdoor Compression.}
For inference on the \ac{DPU}, Vitis AI performs 8-bit quantization on the parameters and subsequently compiles the quantized model into a computation graph using the \ac{XIR}.
This graph can be serialized into and de-serialized from a proprietary \texttt{.xmodel} file after quantization and compilation.
Such a file contains the layers of the model to be executed and the quantized model parameters.
For inference, the compiled file, which also features the \ac{DPU} instructions, is flashed to the device and executed using the Vitis AI Runtime \acs{API}.

We generate a list of differences between the quantized and compiled parameters of the original model and the backdoored one to use them for initialization of trojan \ac{ROM} later on.
To determine these differences, we compare the \texttt{.xmodel} files of both models.
A quantized \texttt{.xmodel} stores the parameters as 16-bit floats in contrast to the compiled file which uses 8-bit fixed-point values.
Furthermore, the compiled file stores the parameters in an order that is optimized for the shared memory layout.
While the quantized parameters can still be read using Xilinx tools, this not possible for a compiled \texttt{.xmodel} file.
By analyzing the file structure, recovering fixed-point positions, and using a fuzzing-based approach, \ie, generating and comparing compiled \texttt{.xmodel} files for user-defined models, we were able to locate the compiled parameters and automate their extraction.

\par\smallskip\textbf{Backdoor Loading.}
Having computed the model differences, we reverse-engineered the order in which the parameters are flashed to shared memory using known test patterns, as this order differs from the one in which the compiled parameters are kept in the \texttt{.xmodel} file.
Finally, we initialized the \ac{ROM} with the manipulated parameters through a bitstream update.

\subsection{Evaluation}
\label{dpu_trojan::sec::eval}
We evaluated our attack on the Xilinx Zynq UltraScale+ MPSoC ZCU104 by running inference on the trojanized \ac{DPU} using the test data from \autoref{dpu_trojan::subsec::ml_eval}. 
We settled for a backdoored VGG-16 model generated using $L^1$ regularization and a trigger size of $50\times50$ pixels.
This setup requires seven weight changes to achieve a trigger \ac{DSR} of $90\%$ before quantization, see \autoref{dpu_trojan::subtab::trigger_size_eval}.

\begin{figure}
    \centering


\definecolor{col1}{HTML}{332288}
\definecolor{col2}{HTML}{117733}
\definecolor{col3}{HTML}{88CCEE}
\definecolor{col4}{HTML}{DDCC77}
\definecolor{col5}{HTML}{882255}

\begin{tikzpicture}

\begin{groupplot}[group style = {group name=gp,group size = 1 by 2,
    horizontal sep = 40pt,
    vertical sep=60pt},
    width = 0.9\linewidth,
    height = 0.5\linewidth
]

\nextgroupplot [
      xlabel={\# Replaced parameters},
      ylabel style={align=center},
      ylabel={Percentage [\%]},
      axis on top,
      grid style={on layer=axis background},
      label style={font=\small},
      ymin=25, ymax=100,
      ytick={25,50,75,100},
      legend style = {column sep = 3pt, legend columns = 1, font=\footnotesize},
      tick label style={font=\small},
      legend cell align={left},
      legend pos=south east,
      enlarge x limits=0,
      enlarge y limits=0,
      grid = both,
      clip mode = individual
    ]
\node[text width=1em,anchor=north west] at (axis cs:50,8) {\subcaption{\label{dpu_trojan::fig::eval}}};
\addplot[color=black, line width=0.3mm] plot [] table [x=name, y=success-rate, col sep=comma]{data/hw-trojan/evaluation-casestudy.csv};
\addlegendentry{Success Rate}

\addplot[color=black, line width=0.3mm, dashed] plot [] table [x=name, y=accuracy, col sep=comma]{data/hw-trojan/evaluation-casestudy.csv};
\addlegendentry{Accuracy}

    \nextgroupplot [
      title = {},
      xlabel={\# Replaced parameters},
      ylabel style={align=center},
      ylabel={Overhead [\%]},
      axis on top,
      grid style={on layer=axis background},
      label style={font=\small},
      ymin=0, ymax=1.5,
      legend style = {column sep = 3pt, legend columns = 1, font=\footnotesize},
      tick label style={font=\small},
      legend cell align={left},
      legend pos=north west,
      enlarge x limits=0,
      enlarge y limits=0,
      grid = both,
      clip mode = individual
    ]

\node[text width=1em,anchor=north west] at (axis cs:50,-0.35) {\subcaption{\label{dpu_trojan::fig::overhead}}};
\addplot[color=black, line width=0.3mm] plot [] table [x=name, y=lut, col sep=comma]{data/hw-trojan/overhead-casestudy.csv};
\addlegendentry{LUTs}

\addplot[color=black, line width=0.3mm, dashed] plot [] table [x=name, y=ff, col sep=comma]{data/hw-trojan/overhead-casestudy.csv};
\addlegendentry{FFs}

\addplot[color=black, line width=0.3mm, dotted] plot [] table [x=name, y=lutram, col sep=comma]{data/hw-trojan/overhead-casestudy.csv};
\addlegendentry{LUT-RAM}
    
\end{groupplot}
\end{tikzpicture}
    \caption{(a) Success rate and test accuracy for backdoored variants of the traffic sign recognition model when being executed on the Xilinx Vitis AI \acs{DPU}. (b) Hardware \trojan overhead required to realize the respective number of weight replacements. 
    The original \ac{DPU} utilizes \numprint{37379} \acsp{LUT}, \numprint{6440} LUT-RAM, and \numprint{90309} \acsp{FF}.}
    \label{fig:enter-label}
\end{figure}

\autoref{dpu_trojan::fig::eval} shows the trigger success rate and test accuracy of the backdoor after quantization.
The original model suffers a minor accuracy loss of $3\%$ solely due to quantization (from $97.43\%$ to $94.49\%$).
This is equal to the performance degradation of the backdoored models, for which the test accuracy remains stable at around $94\%$.
As quantization causes deterioration of the trigger success rate compared to the $90\%$ \ac{DSR} achieved with seven parameter changes before, we gradually increase the number of changes up to $100$.
The success rate converges to 83\% while reaching the final plateau after $40$ changes.

\autoref{dpu_trojan::fig::overhead} depicts the hardware overhead in the number of \acsp{LUT}, \acsp{FF}, and LUT-RAM being used for a varying number of replaced parameters.
The more parameters we replace, the more memory lines must be kept in the trojan \ac{ROM}.
If manipulations spread across multiple load instructions, the additions to the memory reader \ac{FSM} become more complex as the trojan then needs to check against multiple addresses, thus requiring more resources.
To cater for potential model updates and allow for larger backdoors, sufficient \ac{ROM} should be provisioned during trojan insertion.
Here, our trojan implementation causes a total hardware overhead below 1\% and fits the target device.
In the absence of a golden model, this results in a stealthy trojan implementation as no unreasonable amount of resources is required to mount the manipulation.
No delay in terms of clock cycles is added to the implementation, hence inference times are equal to the original \ac{DPU}.
Based on these results, we argue that 30 weight changes resulting in a success rate of $78.15\%$ are a good trade-off to cause significant harm at little overhead.
\section{Discussion}
\label{dpu_trojan::sec::discussion}
In this section, we discuss the implications and countermeasures of the presented attack from both the hardware and machine learning perspectives.
\subsection{Implications}
\label{dpu_trojan::subsec::discussion:implications}

\par\smallskip\textbf{Hardware Acceleration.} 
By realizing a backdoor that is added to a learning model strictly within the hardware, we bypass all software and model integrity checks aimed at ensuring valid predictions. 
Our work thus demonstrates that the hardware used for machine-learning acceleration cannot be blindly trusted and must undergo the same scrutiny as the software and learning model to ensure correct and trustworthy operation. 
In safety-critical scenarios, the use of closed-source third-party accelerators for machine learning must be questioned, as they pose a potential security risk.


\par\smallskip\textbf{\acs{ASIC} vs. \acs{FPGA} Deployment.}
Our case study targets an \ac{FPGA} accelerator.
Going beyond our attacker model, \acp{FPGA} also allow for a trojan to be injected in-field. 
Given access to the bitstream, an adversary could manipulate the hardware implementation even after deployment.
Although altering bitstreams is tedious, it is well-understood~\cite{DBLP:conf/date/PhamHK17,DBLP:conf/woot/KatariaHPC19,DBLP:conf/fpga/NoteR08,DBLP:conf/aspdac/EnderSWWKP19} and certainly viable for powerful adversaries. 
While bitstream protection schemes exist, they are difficult to implement and apply correctly~\cite{DBLP:conf/uss/Ender0P20,DBLP:conf/fccm/EnderLMP22,DBLP:conf/ccs/MoradiBKP11,DBLP:conf/ctrsa/MoradiKP12,DBLP:conf/cosade/0001S16,DBLP:journals/trets/SwierczynskiMOP15,DBLP:conf/ccs/TajikLSB17}.

We target an \ac{FPGA} due to its accessibility for academic research.
However, our trojan attack carries easily over to \acp{ASIC}. 
For example, Google's TPU~\cite{DBLP:conf/isca/JouppiYPPABBBBB17} features an architecture similar to the Xilinx DPU, which enables the same attack to be applied to their architecture.
Consequently, circuitry for swapping selected weights, as described in \autoref{dpu_trojan::subsec::trojanizing_dpu}, could be added to many \ac{ASIC} accelerators.
Still, in order to be universally usable, programmability with respect to the backdoor parameters is strictly required.

\subsection{Detectability \& Countermeasures}
\label{dpu_trojan::subsec::discussion:countermeasures}

\par\smallskip\textbf{Detectability.}
The trojan is implanted during design or manufacturing, and our hardware manipulation overhead is minimal.
Hence, as discussed in \autoref{dpu_trojan::sec::intro}, the only viable option for trojan detection is to analyze the circuit itself for malicious functionality.
For \acp{FPGA}, this requires tedious reverse engineering of the bitstream format and, crucially, interpretation of whether there are any malicious functions hidden within an unknown architecture.
For \acp{ASIC}, one needs to image the chip layer by layer using a \ac{SEM} and extract a netlist using computer vision, a task that requires highly specialized equipment, skills, and considerable monetary resources.
Even after successful netlist recovery, one again faces the problem of detecting a \trojan within an unknown circuit.
We claim that such efforts are out of reach in practice.
Although nation-states dispose of the resources to conduct such investigations, the required effort does not scale with the number of samples to be tested.

\par\smallskip\textbf{Hardware Countermeasures.}
Two antagonistic approaches could be followed to harden a hardware design against manipulations.
Cryptographic and obfuscation measures can hamper manipulating the \ac{HDL} design. 
This demands a trusted design process, requiring strict access restrictions for the design files, vetting of all involved employees, and verification of design tools. 
Furthermore, this chain of trust must be extended to all third-party \ac{IP} cores.
Another strategy is switching to an open-source approach and ensuring public access to all design sources, allowing for independent verification.
Although both strategies can help mitigate tampering along the supply chain, a trojan can still be inserted during the manufacturing, for example, by replacing the trusted netlist with a trojanized clone. 
Consequently, using hardware accelerators for security-critical machine-learning applications demands a trusted production facility.

\par\smallskip\textbf{Machine Learning Countermeasures.}
Current approaches for detecting machine-learning backdoors~\cite{DBLP:conf/sp/WangYSLVZZ19,DBLP:conf/sp/XuWLBGL21} fail because our attack operates within the hardware accelerator and the outside model remains unchanged. 
Detecting the backdoor during execution~\cite{DBLP:conf/aaai/ChenCBLELMS19,DBLP:conf/nips/Tran0M18,DBLP:conf/acsac/GaoXW0RN19}, \eg, by monitoring neuron activations, may help, but incurs significant overhead and counteracts the purpose of hardware acceleration. 
The decrease in accuracy induced by our backdoor is similar to that of quantization, so the attack cannot be detected from the model's accuracy either.
To detect the malicious behavior, one needs to compare many outputs of the hardware-accelerated model to the original quantized version running in software. 
While this strategy allows for identifying prediction discrepancies, the backdoor and its trigger remain unknown. 
Currently, we lack appropriate methods to identify backdoors with this hybrid form of hardware-software testing.
Finally, to prevent our trojan from activating, one could permute the parameters streamed to the hardware accelerator.
This renders our attack incapable of identifying the correct insertion point for the manipulated parameters during opeation.
However, this approach is not capable of detecting the trojan in the accelerator.

\section{Conclusion}
\label{dpu_trojan::sec::conclusion}
Our work extends the lively front of adversarial machine learning to a new component: hardware acceleration. 
We investigate the threat of hardware trojans for machine learning and present a programmable trojan framework that backdoors a learning model in hardware during inference. 
All manipulations remain within the hardware and no model changes can be observed, defeating existing defenses. 
To realize the trojan, we expand on the concept of minimal backdoors that require very few parameter changes to implant malicious functionality. 
We demonstrate the applicability of our attack by implanting a trojan in an off-the-shelf accelerator from Xilinx.

Despite making strong assumptions on the attacker's capabilities, we expect the required sophistication to be in reach for well-organized adversaries.
Such supply chain attacks have been a serious concern for many years~\cite{force2005high}, resulting in major investments by governments around the world~\cite{euchips2022,uschips2022}. 
Hence, out trojan attack illustrates that hardware should not be blindly trusted and the integrity of machine-learning accelerators needs to be carefully protected and verified, similar to other security-critical components.
We urge manufacturers, \ac{IP} vendors, and system integrators alike to pay close attention to these threats, and call on the research community to develop countermeasures to defend against this class of attacks.
\section*{Acknowledgements}
We gratefully acknowledge funding by Deutsche Forschungsgemeinschaft (DFG, German Research Foundation) under Germany's Excellence Strategy -- EXC 2092 CASA -- 390781972 and the European Research Council (ERC) under the consolidator grant MALFOY (101043410).

\printbibliography

\end{document}